\documentclass[aps,prx,reprint,superscriptaddress,longbibliography]{revtex4-2}

\usepackage[utf8]{inputenc}
\usepackage{graphicx}
\usepackage{bm}
\usepackage{amsmath,amssymb,amsfonts}
\usepackage{siunitx}
\usepackage{booktabs}
\usepackage{color}
\usepackage[T1]{fontenc}
\usepackage{newtxtext}
\usepackage{newtxmath}
\usepackage[hidelinks]{hyperref}
\usepackage[capitalise]{cleveref}
\usepackage{wrapfig}
\usepackage{float}
\usepackage{subcaption}
\usepackage{nth}
\usepackage{microtype}
\usepackage{xspace}
\usepackage{ragged2e}

\newcommand{\htp}{\ensuremath{\mathrm{H}_2^+}\xspace}
\newcommand{\deut}{\ensuremath{\mathrm{D}^+}\xspace}



\begin{document}

\title{From Beam to Bedside: Reinforcing Domestic Supply of $^{99}$Mo/$^{99m}$Tc using Novel High-Current D$^+$ Cyclotrons for $^{99}$Mo Production}

\author{Jarrett Moon}
\affiliation{Massachusetts Institute of Technology, Cambridge, MA 02139, USA}
\author{Daniel Winklehner}
\affiliation{Massachusetts Institute of Technology, Cambridge, MA 02139, USA}
\author{Jose Alonso}
\affiliation{Massachusetts Institute of Technology, Cambridge, MA 02139, USA}
\author{Claire Huchthausen}
\affiliation{Massachusetts Institute of Technology, Cambridge, MA 02139, USA}
\date{\today}
\author{David McClain}
\affiliation{Massachusetts Institute of Technology, Cambridge, MA 02139, USA}
\author{Janet Conrad}
\affiliation{Massachusetts Institute of Technology, Cambridge, MA 02139, USA}

\begin{abstract}
Technetium-99m ($^{99m}$Tc) is essential to more than 16 million diagnostic procedures performed annually in the United States. It is typically acquired on-site from generators containing molybdenum-99, in turn produced at nuclear reactor facilities. This supply chain involves multiple points of vulnerability, which can lead to shortages and delays with potentially negative patient outcomes. We report on the development of a new family of cyclotrons originally designed for the IsoDAR neutrino experiment, capable of operating at much higher current than typical cyclotrons. When operated with deuterons at \SI{1.5}{MeV/amu} and an anticipated continuous beam current of \SI{5}{mA}, simulations project that such a system would yield $\sim 10^{13}$ neutrons/s using a thin beryllium target. This neutron yield is sufficient, in principle, to support $^{99}$Mo production without the use of highly enriched uranium or reliance on foreign reactors. Simulations and conceptual design studies suggest that the system’s beam dynamics could make it a viable pathway toward decentralized, hospital-based isotope generation. The relatively low energy of the deuterons minimizes activation and safety concerns. This work presents the physics motivation, technical design considerations, and projected neutron yields, outlining a pathway from a neutrino-physics prototype to a biomedical isotope production platform. \\

\hspace{3mm}
\noindent\textit{Corresponding Author}: Jarrett Moon. jarrett@mit.edu. 
\end{abstract}

\maketitle

\section{Introduction}

The production of molybdenum-99 ($^{99}$Mo) remains a critical challenge for nuclear medicine. $^{99}$Mo decays to technetium-99m ($^{99m}$Tc), the single most widely used medical radioisotope in the world. In the United States alone, between 15 and 20 million $^{99m}$Tc imaging procedures are performed annually~\cite{NASEM2018_Mo99_Tc99m}. 

Despite this demand, nearly all $^{99}$Mo used in the United States is imported from aging foreign reactors. Reactor shutdowns, transportation delays, and regulatory pressures have made more robust domestic production of $^{99}$Mo an urgent priority~\cite{ballinger2010short}.

Traditionally, $^{99}$Mo has been produced at reactors because of their high flux of low-energy neutrons. Previous accelerator-based approaches have generally been limited by achievable beam current and neutron yield, preventing production at levels comparable to nuclear reactors. Nevertheless, they do offer several attractive advantages over reactors. They are individually much cheaper, much smaller, safer, can be built close to hospitals -- or even on site -- and do not require highly enriched uranium fuel. 

Here, we describe a system to generate $^{99}$Mo using a new generation of high current cyclotrons as drivers for neutron production coupled with a low-enriched uranium target. The details of this target derive from work performed at Argonne National Laboratory where, using a different neutron source, they demonstrated generation of high specific activity $^{99}$Mo with low contamination suitable for use in generators~\cite{ANLNature,ANL99Mo}. 

Our new class of cyclotron retains the advantages of accelerator-based systems while operating at a much higher design current of 10~mA protons, or 5~mA deuterons, reducing the gap in achievable neutron production relative to reactor-based approaches. We report on mature designs for these cyclotrons, present simulations of $^{99}$Mo production, and discuss a path toward commercial scale systems.

\section{Background and Motivation}

$^{99m}$Tc is one of the most important medical radionuclides. It is a metastable isomer of $^{99}$Tc which decays via emission of a 141~keV gamma. This useful energy in conjunction with the relatively short radioactive ($\sim$6 hours) and biological ($\sim$1 day) half-lives make $^{99m}$Tc one of the most effective available isotopes with no good substitute~\cite{SNMMI2018_Mo99}.

While $^{99m}$Tc is the isotope of direct medical use, medical centers typically obtain $^{99m}$Tc on site from a device called a generator. These contain a quantity of $^{99}$Mo which decays into $^{99m}$Tc with a half-life of 66 hrs. These devices enable needed quantities of $^{99m}$Tc to be ``milked'' from the generator as needed while the $^{99}$Mo decay replenishes the supply~\cite{Gupta2025Mo99Tc99mGenerator}. As such, the production of useful quantities of extractable $^{99}$Mo is our primary goal. 

There are several pathways by which $^{99}$Mo can be generated. The strategy that has proved to most efficiently yield readily separable $^{99}$Mo involves production via fission of $^{235}$U which results in $^{99}$Mo $\sim$6\% of the time. We will focus on this pathway hereafter. The fission of $^{235}$U requires a significant flux of thermal neutrons, that is neutrons with energies $\sim$0.025~eV. 

Nuclear reactors provide this, however, using reactors has significant drawbacks. Nuclear reactors are expensive, large, comparatively dangerous, have significant environmental risks, can rarely be built close enough to medical centers to avoid loss-in-transit, and pose nuclear materials proliferation risks. Further, at present there are few U.S. facilities which have the integrated, licensed end-to-end capability (irradiation positions + target handling + hot-cell radiochemistry + waste management + QA/regulatory + logistics) to produce $^{99}$Mo at scale~\cite{NASEM2016_Mo99}.

These risks have concretely manifested themselves more than once with intermittent shortages appearing starting in the late 2000s as foreign reactors began to age and shut down~\cite{Service2011,Grose2009}. There are no mature long-term plans to replace these with alternative reactors either abroad or domestically. These supply chain risks - in conjunction with nonproliferation pressures to move away from highly enriched uranium (HEU) to low enriched uranium (LEU) targets - have made domestic production an urgent national priority. The American Medical Isotopes Production Act of 2012~\cite{IsotopesAct} directed U.S. agencies to establish a reliable, HEU-free domestic supply chain. 

Accelerator-based neutron sources offer a promising route to achieving this goal without reliance on large reactors. The major limitation on accelerator-based production, especially with cyclotrons, has historically been their relatively low neutron yield making isotope production less attractive than reactors despite their other advantages. A higher current cyclotron has the potential to open this bottleneck, allowing systems with all the advantages of accelerator-based production with yields acceptably close to regional scale $^{99}$Mo production needs. 

A new family of high-current cyclotrons has recently been developed for a particle physics experiment called IsoDAR. The IsoDAR (Isotope Decay-At-Rest) experiment is a particle physics experiment designed to study anomalies associated with particles known as neutrinos as well as other rare processes~\cite{winklehnerIsoDARYemilabPreliminaryDesign2025}. In order to obtain the needed statistics, a uniquely intense and compact source of protons is necessary in order to act as a driver for downstream neutrino production. These new high-current cyclotrons were originally developed to deliver this required intensity.

This new family of cyclotrons, detailed further in Section~\ref{sec:CyclotronDeets}, incorporates several novel features that produce an order of magnitude higher currents than traditional cyclotrons. We refer to these cyclotrons as a family rather than a single device because of their flexibility in ion species and final extracted energy. While originally developed to accelerate \htp, any ion with an equal mass to charge ratio can be accelerated with modest modifications including \htp, \deut, or heavier ions in higher charge states. The energy to which the ions are accelerated can vary from approximately 1-100~MeV/amu depending on the size of the device. We term a specific one of these high current cyclotrons operating at XX MeV/amu and accelerating \htp/\deut an HCHC-XX / HCDC-XX.

These compact devices produce multi-milliampere currents. When operated with D$^+$ and paired with a beryllium target this is capable of generating significant neutron fluxes which can be used for $^{99}$Mo production.

\section{Novel High Current Cyclotrons}\label{sec:CyclotronDeets}

Several key advances have been developed and incorporated into traditional cyclotron design, resulting in the HCHC-XX cyclotron family. These advances enable them to reach much higher currents than prior cyclotrons. We first provide an overview of the HCHC-XX cyclotron family with an emphasis on the new features enabling high currents. Significantly more details can be found in IsoDAR's preliminary design report~\cite{winklehnerIsoDARYemilabPreliminaryDesign2025}. 

These new features broadly apply to any HCHC-XX/HCDC-XX, but we will frequently discuss these cyclotrons with reference to their most mature iteration to date, the IsoDAR experiment's HCHC-60 and to its lower energy prototype HCHC-1.5. 

Constructing an HCHC at other energies or constructing a \deut accelerating HCDC-XX requires only modest modifications. We will highlight and discuss these modifications required to operate with deuterons rather than molecular hydrogen. In particular, we will discuss modifying the prototype HCHC-1.5 into an HCDC-1.5.

\subsection{HCHC-XX}

The HCHC-XX is a room-temperature, compact, isochronous (azimuthally varying magnetic field) cyclotron with four hills and four double-gap radio-frequency (RF) cavities. The ions are initially produced using a filament driven multicusp ion source~\cite{winklehnerMIST1MIST2Multicusp2025}. A low energy transport line guides the ions into a radiofrequency quadrupole (RFQ) which is embedded directly into the cyclotron. This RFQ operates at a uniquely low frequency (32.8~MHz) in order to match the cyclotron frequency. To accomplish this low frequency, a split coaxial design is used~\cite{winklehnerHighCurrentH2Beams2021,holtermannTechnicalDesignRFQ2021}.

\begin{figure}[h]
    \centering
    \includegraphics[width=0.8\linewidth]{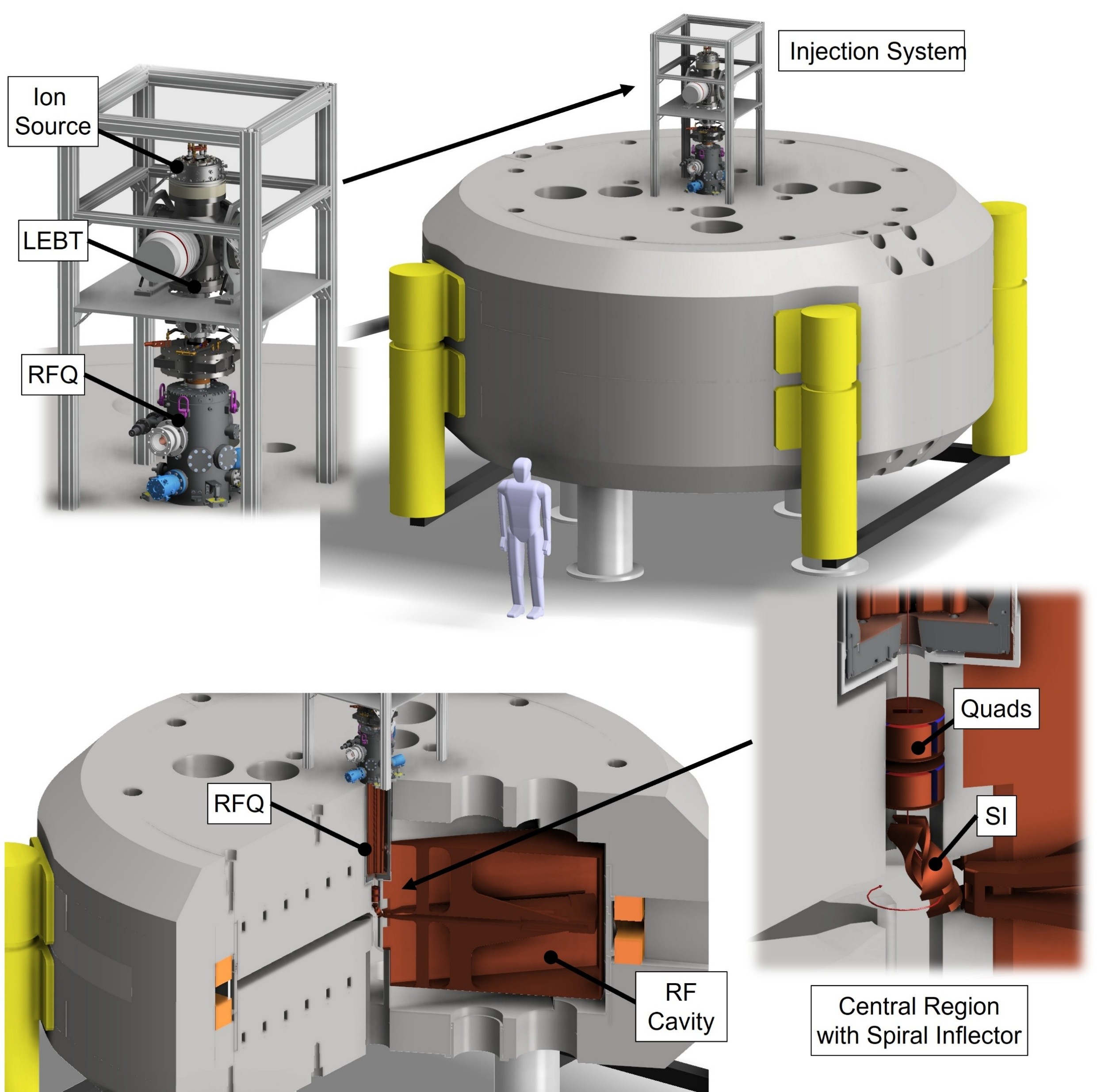}
    \caption{Design overview of an HCHC-60, showing ion source, low energy bream transport (LEBT), radiofrequency quadrupole (RFQ), and cyclotron with spiral inflector (SI) and RF cavities. From~\cite{winklehnerIsoDARYemilabPreliminaryDesign2025}}
    \label{fig:HCHC-60_Overview}
\end{figure}

The ions exit the RFQ and pass through a set of electrostatic focusing quadrupoles which maintain the focusing. They then pass through a spiral inflector, a device comprising a pair of helical electrodes which bend the beam 90 degrees from vertical into the primary accelerating plane of the cyclotron~\cite{winklehnerIsoDARYemilabPreliminaryDesign2025}. These key components are illustrated for an HCHC-60, which is planned to be used for the IsoDAR experiment, in Figure~\ref{fig:HCHC-60_Overview}.

After reaching the desired energy, a resonance is excited using iron bars, which generates enough turn separation to use an electrostatic septum to extract the beam. If \htp is accelerated, a stripper foil in the transport line can remove an electron to convert each \htp to a proton pair, leading to a final doubling of extracted electrical beam current. A non-molecular ion beam does not require this step. In the case of \deut, the extracted beam will impact a neutron-generating beryllium target.

\begin{figure}[h]
    \centering
    \includegraphics[width=1.0\linewidth]{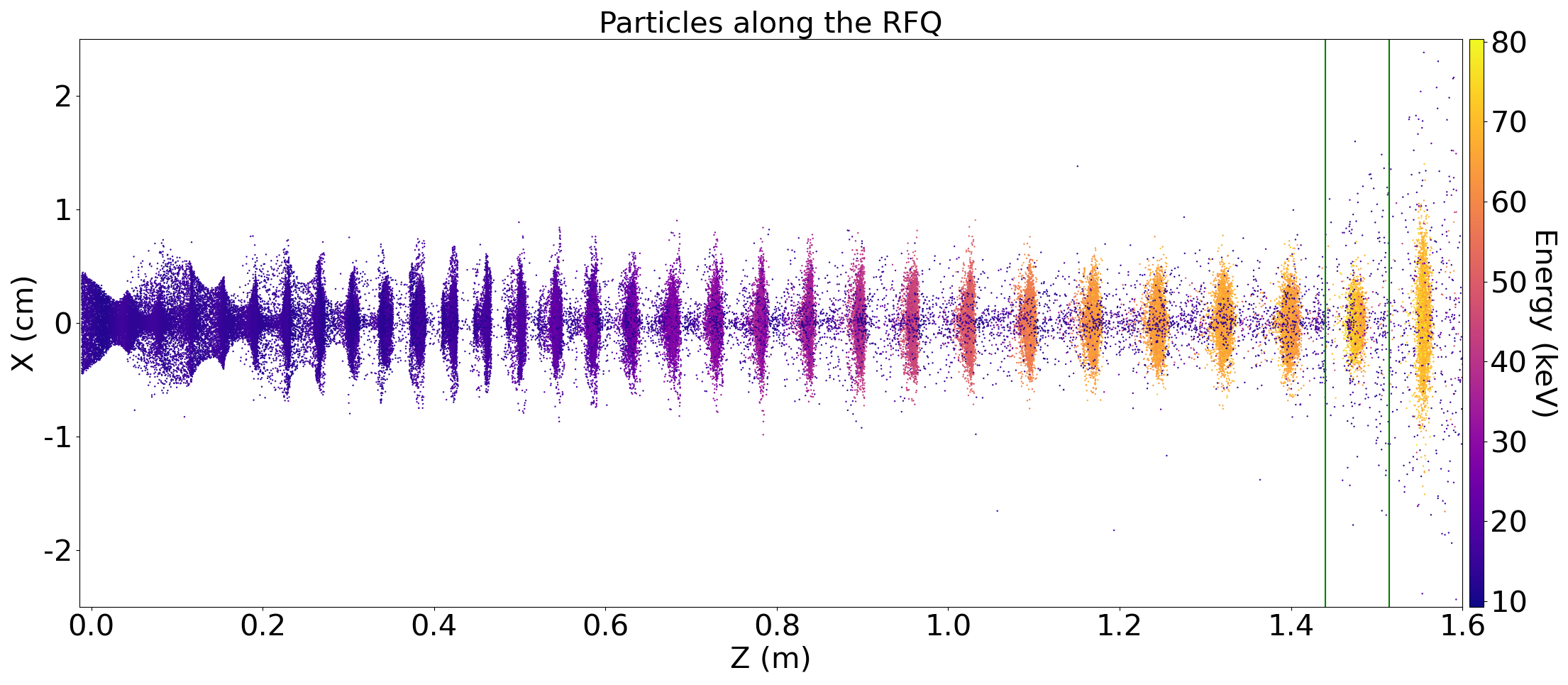}
    \caption{An illustration of the ion beam entering the RFQ where it is alternatively focused in the transverse and vertical planes. At beam exit, corresponding to the bunch between the vertical green lines, we see a clean bunch has formed.}
    \label{fig:RFQPreBunch}
\end{figure}

The HCHC-XX design achieves it's high current by leveraging several key new features:

\begin{enumerate}
    \item \textbf{Accelerate Molecular Hydrogen Ions:} When charged particles are accelerated, they are done so in bunches. A bunch of like-charged particles experiences ``space charge'' effects, that is, they self-repel due to the Coulomb force. This worsens at higher currents. Accelerating \htp rather than protons reduces space charge limitations as, per particle, one negative charge is accelerated in tandem with two positive charges.
    
    \item \textbf{Pre-Bunching:} After ions are generated, they are transported into the RFQ axially embedded in the cyclotron. RFQs provide both acceleration and very efficient longitudinal and transverse focusing. In an HCHC-XX the RFQ primarily acts as a high efficiency buncher. The ions are accelerated only modestly to $\sim$35~keV/amu, but before injection are efficiently bunched as seen in Figure~\ref{fig:RFQPreBunch}. 
    
    The bunch begins to de-focus (grow) almost immediately after exiting the RFQ. To preserve the bunch quality, a pair of electrostatic quadrupole lenses are placed directly after the RFQ exit. The spiral inflector is also a novel design which incorporates geometric modifications that introduce quadrupole moments directly into the electrodes~\cite{barnardLongitudinalVerticalFocusing2021,toprekTheoryCentralIon2000,winklehner_realistic_2017}.

    This entire system provides extremely efficient pre-bunching and focus preservation right until the moment of injection into the central plane of the cyclotron.

    \item \textbf{Vortex Motion:}  Vortex motion is a stabilizing collective multi-particle effect that can occur in high-current beams in cyclotrons. Vortex motion was first noted experimentally at the Paul Scherrer Institute in the PSI Injector II cyclotron~\cite{stetson:vortex,baumgartenTransverselongitudinalCouplingSpace2011}. 
    
    The electric self-fields exert a force directed radially outwards on the ions in the bunch. This radial force in conjunction with the external forces from the cyclotron magnetic field lead to a ``curling up'' of the beam in radial-longitudinal space (looking onto the bunch from the top), generating an almost round beam shape. Vortex motion stabilizes these high density beams, enabling them to remain small until extraction. The HCHC-XX family is the first to be designed with vortex motion in mind and thus has been designed to maximally leverage its ability to stabilize high current bunches. High-fidelity particle-in-cell (PIC) simulations using the well-established code OPAL~\cite{adelmannOPALVersatileTool2019} have verified vortex motion within the present HCHC-XX design.
    
\end{enumerate}

The individual components of an HCHC-XX have all been extensively simulated, including FEA/FEM modeling during the technical design phases. We are now preparing a prototype experiment (an HCHC-1.5). The experiment comes in three phases: 

\begin{itemize}
    \item Ion source: Constructed with preliminary commissioning tests completed and additional tests ongoing.
    \item RFQ: Delivery from vendor (Summer 2026)
    \item Cyclotron: Vendor under selection (Spring 2026) with anticipated 2027 delivery.
\end{itemize}

As none of features unique to an HCHC-XX occur after acceleration to  $\sim$1.5~MeV/amu, a low energy prototype is able to perform all needed tests.

\subsection{HCDC-1.5}
The HCHC-XX cyclotrons were originally envisioned to accelerate \htp. Deuterons, however, offer better neutron production prospects. So we consider modifying the HCHC-1.5 to an HCDC-1.5. Deuterons have a q/m less than 0.1\% different from \htp making them almost automatically suitable, but several small changes will be required to existing designs.

\begin{enumerate}
    \item \textbf{Ion source conversion:} The present ion source has its focusing lenses and plasma generation parameters optimized for $H_2$ gas rather than $D_2$ gas. Mature procedures using the ion source and plasma boundary simulation code IBSimu~\cite{kalvas:ibsimu} have been successfully used to develop the extraction system (electrostatic focusing lenses for initial beam shaping) for \htp and will readily be used for \deut modification.
    
    \item \textbf{Collimator replacement:} A series of low energy collimators, distributed around the first few turns, has been identified and optimized for IsoDAR's HCHC-60, which encourages tight particle bunching and helps permit effective extraction.
    
    The placement of these collimators will require adjustment. Modification will be required because of small differences in acceleration behavior of \deut vs \htp. Further, these collimators are currently optimized for the HCHC-60 which requires significant bunch tightening to allow extraction at such high energies. A lower-energy HCDC will not require such aggressive collimation to permit extraction, thus allowing us to place the collimators at lower energies, mitigating activation concerns.

    This adjustment will be done using the same mature procedure as was used for \htp. The OPAL code~\cite{adelmannOPALVersatileTool2019} is used to simulate acceleration while CERN's FLUKA particle simulation code~\cite{Battistoni2016,Ahdida2022} is used to model collimator activation. 

    \item \textbf{Extraction system.} The electrostatic septum-based extraction system described previously has been developed for IsoDAR's HCHC-60. This extraction system must be adapted for an HCDC-1.5 to enable guiding the \deut beam onto an external target to generate neutrons.
    
\end{enumerate}

The HCHC-1.5 prototype was designed and is being built with these modifications in mind. 

\section{Neutron Production \& Mo-99 Generation}

\subsection{HCDC-1.5 + Beryllium Target}

\begin{figure}[ht]
\centering
\includegraphics[width=0.5\textwidth]{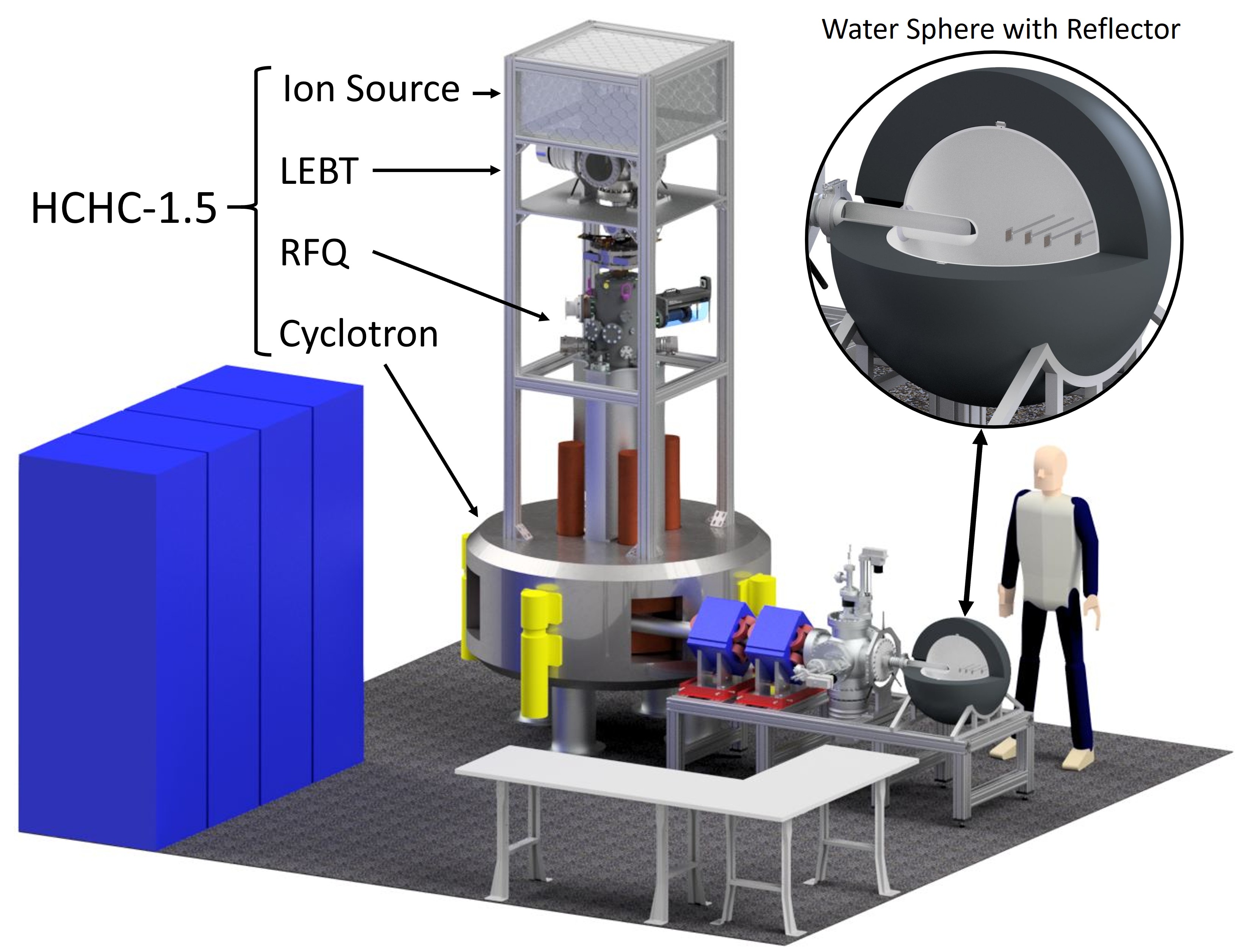}
\caption{Conceptual layout of the HCDC-1.5 with extraction line and thermalization sphere.}
\label{fig:CAD}
\end{figure}

The eventual production of $^{99}$Mo from $^{235}$U fission requires a significant flux of thermal neutrons. This is accomplished by extracting the beam of 1.5~MeV/amu deuterons from the modified HCDC-1.5 and directing it onto a thin beryllium target. This setup is illustrated in Figure~\ref{fig:CAD}

After extraction from the cyclotron, the \deut beam passes down an evacuated beam pipe which is surrounded by a pair of quadrupole magnets which focus the D$^+$ onto the beryllium target. The target itself is a beryllium ellipsoid affixed to the end of the beam pipe. The elliptical moments added to the target are optimized such that the off center portions of the incident \deut beam see an average target thickness similar to those on center. The beam impacts the concave interior surface and traverses a layer of beryllium on average $\sim$50~\si{\micro\meter} thick. 

Beryllium is chosen because, at these relatively low deuteron energies, low-Z targets are superior to high-Z targets because the dominant neutron production pathways have threshold Q-values that suppress conversion. For low-energy deuterons, breakup and stripping reactions -- reactions in which the deuteron sheds its loosely bound neutron -- occur efficiently in the field of light nuclei which have low Coulomb barriers and favorable (d,n) channels with low reaction Q-values. Heavy targets, by contrast, have high Coulomb barriers that suppress the interaction probability at these energies. Their neutron emission channels typically require significantly higher incident energies than present in the HCDC-1.5. As a result, a $\sim$3~MeV D$^+$ on Be produces orders-of-magnitude more neutrons per particle than high-Z materials more often seen in higher energy spallation targets.

\begin{figure}[h!]
\centering
\includegraphics[width=0.5\textwidth]{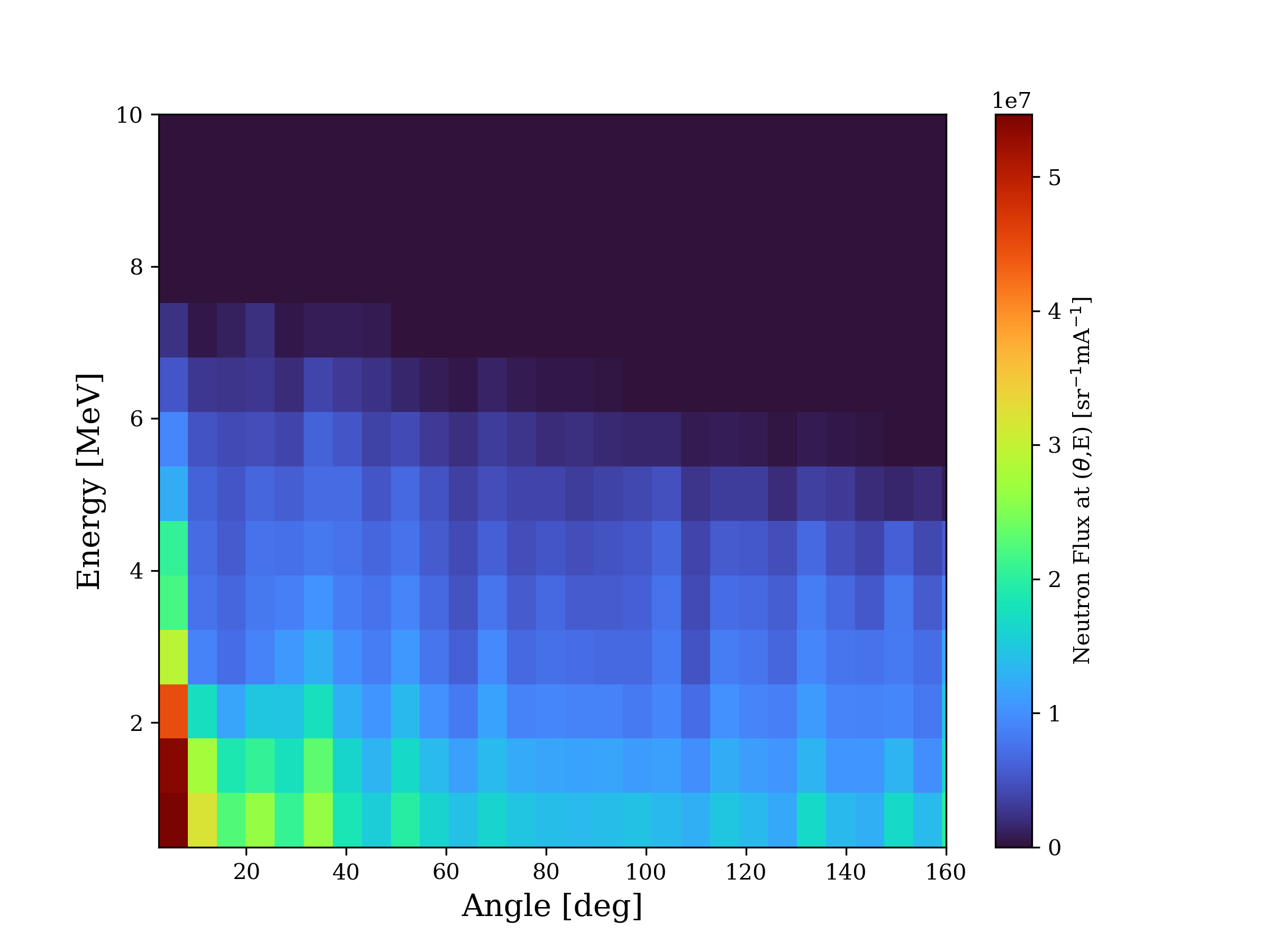}
\caption{Simulated neutron energy and angular distribution for a \SI{1.5}{MeV/amu} D$^+$ beam on a Be target.}
\label{fig:NeutronEvsAng}
\end{figure}

The resulting neutron spectrum emerging from the beryllium target is forward-biased and peaks at \SI{700}{keV}. On average each incident \deut yields 2.8$\times$10$^{-4}$ neutrons. This corresponds to 1.8$\times$10$^{12}$~n/mA$_{\deut}$ or $\sim$9$\times$10$^{12}$~n/s for an HCDC-1.5 running at full 5~mA current. The distribution of neutrons emerging from the target are illustrated in Figure~\ref{fig:NeutronEvsAng}. 

\subsection{A Thermalizing Target}

A significant fraction of the neutrons emerging from the beryllium target are too high-energy to induce significant $^{235}$U fission. Our setup, leveraging the work done at Argonne~\cite{ANLNature,ANL99Mo}, merges the fission target and thermalizing medium into a single structure. The thermalization medium, water, doubles as a solvent for a uranyl-sulfate salt acting as the fission target. This arrangement produces a compact, self-thermalizing fission target which, as we will discuss further in~\ref{sec:UsulfTarget}, permits recycling and product extraction.

The beryllium target at the end of the beam extraction pipe passes through a flange on the side of an aluminum spherical vessel which is filled with a saturated solution of dissolved uranyl-sulfate. The uranium within the uranyl-sulfate is itself enriched to 19.75\% $^{235}$U, below regulatory thresholds for highly enriched uranium and compliant with the goals set out in the American Medical Isotopes Production Act~\cite{IsotopesAct}.

Unlike conventional neutron production targets -- in which neutrons are produced and subsequently moderated in a separate assembly -- this configuration directly embeds the target nucleus into the moderating material. Water provides an efficient hydrogenous moderator, slowing the initially fast neutron spectrum to the thermal range.  By dissolving the uranium directly into this moderating volume, we ensure that a large fraction of the thermalized neutrons will encounter $^{235}$U.

This approach also simplifies heat removal and geometrical design. Because water simultaneously functions as moderator, coolant, and solvent, deposited beam power is efficiently distributed through a high-heat-capacity fluid. To maximize the efficiency of the target as well as to minimize neutron leakage into the environment, a 10~cm thick shell of neutron reflective graphite encases the inner water filled aluminum sphere. A conceptual schematic of the entire HCDC-1.5 + extraction + thermalizing target array is shown in Figure~\ref{fig:CAD}

While not necessary for a production version of this device, the planned $^{99}$Mo production prototype based on the HCDC-1.5 cyclotron design will also embed a series of neutron counting foils inside the target volume to validate the simulation results. The foils~\cite{shieldwerx_SWX1553_ActivationFoils} will provide measurements from 0.0025 eV - 13.5 MeV, fully covering our region of interest.

\subsection{Uranyl-Sulfate Target}\label{sec:UsulfTarget}

Beyond the yield and compactness benefits provided by the self-thermalizing target, we have also chosen uranyl-sulfate building on work performed at Argonne National Laboratory, which demonstrated that it provides a target amenable to extraction and recycling~\cite{ANLNature,ANL99Mo}. Their work used a different neutron source, but demonstrates that uranyl-sulfate dissolved in water is an effective fission target as validated by irradiation of uranyl-sulfate solutions. 

They find that dissolution and initial handling steps are dramatically simplified relative to solid U$_3$O$_8$ targets. The authors show that after irradiation the uranyl-sulfate matrix forms a clean, homogeneous aqueous feed which can be readily drained as needed and processed downstream with chemical separations without the need for powder dissolution, filtration, or mechanical disassembly. 

This design also makes the target intrinsically compatible with recycling, since the depleted uranium solution can be restored and reused after $^{99}$Mo extraction. As such, the study provides strong operational evidence that uranyl-sulfate in water is a robust “solution target” suitable for repeated irradiation–processing cycles.

A major conclusion of their work is the demonstration that $^{99}$Mo can be efficiently and selectively recovered from the irradiated solution, achieving the high specific activity needed for medical-grade $^{99m}$Tc generators. 
Their work suggests that aqueous uranyl-sulfate targets are not only workable, but a potentially superior choice of fission target isotope production, extraction, and iterative reuse.

\subsection{Simulated $^{99}$Mo Production}

Simulation of $^{99}$Mo production was performed in several steps. OPAL~\cite{adelmannOPALVersatileTool2019} was used to simulate the acceleration of \deut to extraction after which the beam was imported into CERN's FLUKA software~\cite{Battistoni2016,Ahdida2022} which provided high fidelity simulations of the neutron yield. 

FLUKA models the entire chain of particle transport, from the incident deuteron beam through neutron production in the beryllium and finally the neutron-driven fission reactions inside the uranyl-sulfate solution. No restriction on process number is enforced, i.e. the entire chain of reactions including secondary, tertiary, etc. neutron production and fission is simulated. 

The simulation defines the deuteron source term with appropriate energy, emittance, and spatial characteristics as informed by the endpoint of the OPAL simulation. From there, FLUKA’s hadronic interaction models track deuteron slowing, breakup, and (d,n) reactions within the beryllium. These models generate the full fast-neutron spectrum emerging from the converter, including angular distributions and any secondary charged particles. The neutron field is then propagated into the surrounding  water-based uranyl-sulfate. FLUKA handles scattering, reflection, moderation, and capture processes. 

By coupling these stages within a single, self-consistent transport environment, FLUKA provides a high fidelity estimate for the quantity of $^{99}$Mo fission fragments produced. 

We assume 75\% beam-on time, in line with currently expected maintenance and operation requirements. With this, our  simulations indicate an annual $^{99}$Mo production of $\sim$25~TBq ($\sim$670~Ci) at an HCDC-1.5's full operational 5~mA.

Using the process developed by Brown et al., extraction efficiencies of $^{99}$Mo above 95\% are  achievable~\cite{ANLNature}. Following extraction, the primary remaining losses are associated with time, that is, with the intrinsic decay of the $^{99}$Mo. In typical existing commercial supply chains, the separation, generator manufacture, and shipping time leads to an overall production-to-hospital time on the order of 2-3 days~\cite{nas_mo99_2016,cutler2014diversification}.

Including these losses, this corresponds to delivered $^{99}$Mo-in-generator activity of $\sim$11-15~TBq/year ($\sim$300-400~Ci/year). These values are commensurate with the requirements of a typical mid-sized hospital~\cite{NAS2009}. Note that delays between production and delivery are one of the more significant sources of loss and can plausibly be reduced for smaller systems such as this that may be built more locally.

The production of isotopes in addition to $^{99}$Mo is also simulated. These contaminants predominantly arise from the fission of $^{235}$U into fragments other than $^{99}$Mo. As such, most of these contaminants are irreducible, if anticipated, due to the fundamentally stochastic nature of uranium fission. These do not pose an issue for purification and $^{99}$Mo recovery~\cite{ANL99Mo}, but any prospective infrastructure must allow for this additional source of radiation.

\section{Practical Comparisons}
An HCDC-1.5-driven system is able to reach regional scale needs. Beyond offering sufficient supply, an accelerator-driven system offers concrete benefits vs alternatives. Here we will consider the performance of an HCDC-1.5 system relative to other options including reactor-based production and alternative accelerators on the market.

\begin{center}
\textbf{Size}
\end{center} 

The physical footprint of an HCDC-1.5–driven system is very modest compared to reactor facilities and comparable to that of existing commercial medical accelerator facilities. The complete HCDC assembly requires $\sim$25~m$^2$ with a height of $\sim$5 m, placing it well within the scale of routinely deployed isotope-production accelerators. Modern commercial and hospital-based isotope cyclotrons are designed for installation in radiopharmacy or industrial environments, with typical footprints under 100~m$^2$~\cite{IBA_RadiopharmacyFootprint, HospitalCyclotronFacility}.

LINAC-based isotope production systems similarly require shielded target vaults and associated radiochemistry infrastructure, with facility size dominated by shielding, power delivery, cooling, and isotope handling rather than the accelerator itself \cite{PhotonuclearIsotopeLinac}. 

From a size perspective, an HCDC-1.5 based system is thus a significant improvement over reactor facilities and falls squarely within the established building scales currently used for commercial radioisotope generation.

\begin{center}
\textbf{Cost}
\end{center}
An HCDC-1.5–driven system will be comparably affordable. The projected cost of a first-of-kind prototype is $\sim$\$1.5–2.0 million, which provides a realistic upper bound on anticipated commercial system cost. This places the HCDC-1.5 well below the scale of reactor-based isotope production, where even small research reactors typically require investments of tens of millions of dollars and power-generation–scale reactors run into the billions.

The highest-current commercial isotope cyclotrons presently available operate at proton currents on the order of $\sim$1~mA. Demonstrations up to 1.6~mA on 30~MeV machines have been reported at ASCI's TR-30~\cite{Sabaiduc2007TR30}. Other cutting edge machines such as IBA's Cyclone 70 have reported currents of 750~\si{\micro\ampere}~\cite{IBA_Cyclone70_Brochure}. 

Although detailed pricing for such systems is not generally published, they are widely understood to reside in the multi-million-dollar range. As compared to other leading options, the HCDC-1.5 represents a significant leap from a current-per-dollar perspective.

We note that mA-scale deuteron accelerators do exist in some specialized systems, but almost exclusively outside the energy and application regime relevant to isotope production. Compact DD and DT neutron generators routinely operate with deuteron beam currents in the $\sim$1-10mA range, but at beam energies in the keV rather than multi-MeV scale required by isotope systems~\cite{DTNeutronGenerators,CompactDDGenerator}. 

If we are willing to consider larger systems, mA scale deuteron beams at tens of MeV have been constructed at large, facility-scale LINAC projects, such as SARAF and IFMIF/LIPAc. However, these are both prohibitively large and involve significant additional infrastructure. Accordingly they have not been used as commercial isotope-production systems~\cite{SARAF,IFMIF_LIPAc}. 

Further, even while typically operating at lower currents, these systems do not offer a cost advantage over an HCDC-1.5 system, still requiring multi-million dollar investments~\cite{DTGeneratorCostReview}.

\begin{center}
\textbf{Operational Stability}
\end{center}

Stability is an important factor as it directly impacts the overall production rate. The 75\% up-time we are using for calculating the $^{99}$Mo production rate is a conservative estimate based on the technical specifications of the IBA Cyclone series of cyclotrons~\cite{CycloneIKON}, operational experience with the PSI high-intensity proton accelerator complex (HIPA)~\cite{grillenbergerStatusFurtherDevelopment2013}, and a substantial additional margin of error, due to the unprecedentedly high beam current we are aiming for. Cyclotron down-time comprises scheduled maintenance, unscheduled maintenance (i.e. long electrical trips), and short trips (resolved in under 5~min.). Requirements for commercial cyclotrons for radioisotope production are typically $>95$\% up-time and less than 5 trips/day. PSI reports statistics on their HIPA facility in annual reports and as conference proceedings and peer-reviewed papers (e.g., Ref.~\cite{grillenbergerHighIntensityProton2021a}). HIPA routinely has above 85\% up-time, with the ring cyclotron and the target actually being the cause for the majority of the short trips. From these experiences, we expect a 85\% up-time is possible after an initial commissioning phase, but remove another 10\% for safety.

In contrast, reactor based production is subject to a different, and more importantly, a less predictable set of stability constraints. While ostensible up-time can be high during routine operation, effective up-time for isotope production is tightly coupled to reactor operating schedules, refueling, and regulatory requirements. Scheduled outages are typically long and unavoidable, and unplanned shutdowns often lead to extended interruptions rather than the short, recoverable trips typical of accelerators. As a result, $^{99}$Mo production tends to be episodic, with prolonged periods of non-production. The fundamental nature of radioactive isotopes makes stockpiling infeasible.

Reactor availability is also increasingly influenced by non-technical factors: aging infrastructure, fuel supply chain constraints, and increasing regulatory complexity. Many reactors historically responsible for global $^{99}$Mo production are operating well beyond their design lifetimes, leading to increasingly onerous and time consuming safety review processes. These factors complicate production forecasting and have contributed to repeated supply disruptions.

\begin{center}
\textbf{Supply Chain Stability}
\end{center}

Reactor-based  production is intrinsically centralized with only a few large facilities supplying the global market through relatively complex and vulnerable international logistics. Outages at a single major reactor can significantly curtail the global supply. This, compounded with long transport distances and international logistics, amplifies disruption risk. These vulnerabilities have been well documented during repeated supply crises over the past two decades, which were driven, not by demand fluctuations, but by unexpected reactor outages or extended maintenance periods \cite{IAEA_Mo99_Supply,OECD_Mo99_Crisis}.  Accelerator-based production enables a fundamentally different supply chain and can mitigate or eliminate most of the vulnerable points in existing supply chains. 

Cyclotron-driven systems such as HCDC-1.5 can be deployed in a modular, distributed manner. They can be closer to end users, reducing reliance on long-distance transport. The modularity and affordability of the individual units further enables granular adaptation to local supply and demand. In such a distributed model, the loss of an individual unit reduces total capacity incrementally rather than catastrophically.

This shift from centralized to distributed production has been explicitly identified by international agencies as a key pathway for improving long-term resilience of medical isotope supply chains \cite{IAEA_Accelerator_Production,OECD_FullCostRecovery}. In this context, the HCDC-1.5 system does not merely substitute for reactor capacity, but supports a structurally more robust supply model that is better aligned with the time-sensitive nature of medical isotope delivery.

\section{Safety}

Achieving 5~mA deuteron current presents unique challenges due to activation and heat deposition. Commercial D$^+$ cyclotrons typically operate below \SI{0.1}{mA}, limited primarily by activation from beam losses. An HCDC-1.5 mitigates these risks through three key strategies: it operates at \SI{3}{MeV}, below most activation thresholds; employs strategic placement of beam collimation to confine losses; and utilizes a beam chopper to enable low-duty-cycle commissioning and reduce average activation during tuning. These features collectively reduce activation risk and maintenance downtime while preserving high instantaneous current for neutron production. 

We break the radiation safety into two components: the HCDC-1.5 system itself, including any radiation arising from interactions of the beam with the device, and total radiation including that induced in the target. We isolate the device for comparison with other machines. The activity induced in the target will dominate, but will be in common for any system using uranium fission at these generation levels. 

Prospective worker dose was estimated by taking the maximum simulated beam-on ambient dose and weighting by the most recent conversion coefficients~\cite{ICRP116}. At maximum this yields an equivalent dose of 32~\si{\micro \sievert}/hr. This is not negligible but is well within standard occupational exposures with proper protocols.

The total activity, including both the $^{99}$Mo and contaminants, induced in the target amounts to $\sim$3~TBq per operational day, significantly dominant to radiation from the machine itself. While the cumulative activity in the target is quite high, these inventories are generally managed within shielded systems and controlled areas. Specific shielding requirements for this target will need to be developed and are beyond the scope of this work. We note for now that these activity levels are not outside those handled at radioisotope production facilities at present and do not present a barrier to implementation.

\section{Conclusion}

Accelerator-based production of isotopes is not a new idea, but the currents at which cyclotrons can operate has historically been limited, making them less attractive and uneconomical for neutron-based production due to low yields. The HCHC-XX / HCDC-XX cyclotron family developed for the IsoDAR experiment represents a leap forward in opening this bottleneck with its order of magnitude higher achievable currents.

Simulations have demonstrated that conversion from an $\htp$ accelerating HCHC-1.5 to D$^+$ accelerating HCDC-1.5 is straightforward. With a \SI{1.5}{MeV/amu} D$^+$ beam on a thin beryllium target, the resulting neutron yield reaches nearly $\sim10^{13}$~neutrons/s. These neutrons, when directed onto a self-thermalizing aqueous uranyl-sulfate target, can produce $^{99}$Mo at rates commensurate with the needs of a medium-sized hospital. Previous work has already established the feasibility of extracting this $^{99}$Mo for use and recycling the target material.

Future steps include experimental measurement of neutron yields and direct comparison to LEU target performance models. If validated, the HCDC-1.5 design could be adopted by commercial cyclotron manufacturers, fostering a new generation of compact accelerators for medical isotope production.

When considered in the context of supply chain stability, economics, safety, and patient access, an HCDC accelerator-driven production system provides a promising new approach that is amenable to smaller scale commercialization while also pushing the frontiers of compact accelerator design.

\section{Acknowledgements}

This work was supported by the U.S. National Science Foundation under award PHY-2411745. We also thank the Heising-Simons Foundation for their support.

\bibliography{bibliography_fixed_v2}

\end{document}